\documentclass[aps,prl,preprint,groupedaddress,showpacs]{revtex4-1}

\usepackage{graphicx}

\begin{document}


\title{Numerical evidence of drift term in two-dimensional point vortex system at negative absolute 
temperature}


\author{Yuichi Yatsuyanagi}
\affiliation{Faculty of Eduation, Shizuoka University, Suruga-Ku, Shizuoka 422-8529, Japan}

\author{Tadatsugu Hatori}
\affiliation{National Institute for Fusion Science, Toki, Gifu 509-5292, Japan}


\date{\today}

\newcommand{\wz}{\omega}
\renewcommand{\vr}{\vec{r}}
\newcommand{\vu}{\vec{u}}
\renewcommand{\vec}[1]{\mbox{\boldmath$#1$}}
\newcommand{\grad}{\nabla}
\renewcommand{\div}{\nabla \cdot}

\begin{abstract}
The drift term appearing in an anaylitically obtained kinetic equation 
for a point vortex system is evidenced numerically.
It is revealed that the local temperature in a region where the vortices are 
frequently transported by the diffusion and the drift terms characterizes 
system temperature and its sign is definitely negative.
Simulation results clearly show a transport process of the vortices 
by the diffusion term (outside the clumps) and the drift term (inside the clumps), 
which gives a key mechanism of the self-organization, i.e., condensation of the
same-sign vortices.
\end{abstract}


\maketitle


Large-scale structure formation and self-organization have attracted much attention
in the context of two-dimensional (2D) turbulence.
To describe the energy inverse cascade in a 2D point vortex system,
Onsager introduced a concept of negative absolute temperature 
due to a limited phase-space volume \cite{Onsager}.
The negative temperature state in the point vortex system is numerically evidenced first 
by Joyce and Montgomery \cite{Joyce}, and later by Yatsuyanagi \cite{Yatsuyanagi2005}.
However, details how the point vortex system relaxes to a thermal equilibrium state remain unclear.
Thus, kinetic Eqs. for the point vortex system have been developed under several 
assumptions \cite{Chavanis2001,Chavanis2007,Chavanis2008,Chavanis2012,DubinJin2001,Dubin2003,Yatsuyanagi2015}.
In this context, we have analytically demonstrated a crucial role of the drift term in the
self-organization of the point vortex system at negative temperature \cite{Yatsuyanagi2015b}.
In this paper, we will evidence the role of the drift term numerically.

\begin{figure}[ht]
  \resizebox{6cm}{!}{\includegraphics{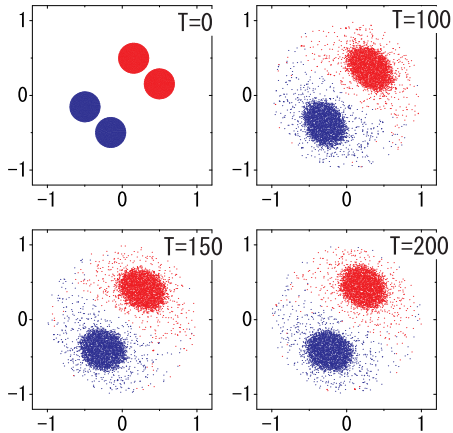}}
  \caption{A typical time evolution of the point vortex system at negative temperature is shown. 
  The number of the vortices is $N=6940$.
  Typical self-rotation time of a clump is $2.4$.}
  \label{fig:fig01}
\end{figure}

Let us consider a point vortex system consisting of $N/2$ positive 
and $N/2$ negative vortices confined in a circular area with radius $R$
\begin{eqnarray}
	\wz(\vr,t) & = & \sum_{i=1}^N \Omega_i \delta(\vr - \vr_i)\nonumber\\
	& = & \sum_{i=1}^{N/2} \Omega \delta(\vr- \vr_i) - \sum_{i=(N/2)+1}^{N} \Omega \delta(\vr- \vr_i)\nonumber\\
	& \equiv & \wz_+(\vr,t) + \wz_-(\vr,t)
\end{eqnarray}
where $\Omega$ is a positive constant, 
$\delta(\vr)$ is the Dirac delta function in 2D
and $\vec{r}_i  = (x_i, y_i)$ is the position vector of the $i$-th point vortex.
Motions of the vortices are traced by the following equations of motion
with the discretized Biot-Savart integral
\begin{equation}
	\Omega_i \frac{dx_i}{dt} = \frac{\partial H}{\partial y_i}, \quad
	\Omega_i \frac{dy_i}{dt} = -\frac{\partial H}{\partial x_i} \label{eqn:canonical}
\end{equation}
where $H$ is the system energy given by
\begin{eqnarray}
	H & = & \sum_i^{N} H_i, \\
	H_i & \equiv & \frac{1}{2} \Omega_i \psi_i, \label{eqn:Hi}\\
	\psi_i & \equiv & \psi(\vec{r}_i) \nonumber\\
		& = & -\frac{1}{2\pi} \sum_{j \neq i}^{N} \Omega_j \log |\vec{r}_i - \vec{r}_j| 
			+ \frac{1}{2\pi} \sum_j^{N}       \Omega_j \log |\vec{r}_i - \bar{\vec{r}}_j| \nonumber\\
		& & - \frac{1}{2\pi} \sum_j^{N}           \Omega_j \log \frac{R}{|\vec{r}_j|}. \label{eqn:psi_i}
\end{eqnarray}
The notation $H_i$ corresponds to the energy possessed by the $i$-th point vortex.
The effect of the circular wall is introduced by the image vortices at
 $\bar{\vec{r}}_i = R^2\vr_i/|\vr_i|^2$.
A typical time evolution of the point vortex system is shown in Fig. \ref{fig:fig01}.
Initial 4 clumps are destroyed and 2 clumps are produced rapidly 
by the violent relaxation \cite{LyndenBell1967}.
Each clump at $T=100$ consists of exclusively single-sign vortices.
After the violent relaxation, the system reaches a slow relaxation
phase, 
during which 2 clumps are stuck on almost the same positions.

\begin{figure}[ht]
  \resizebox{8cm}{!}{\includegraphics{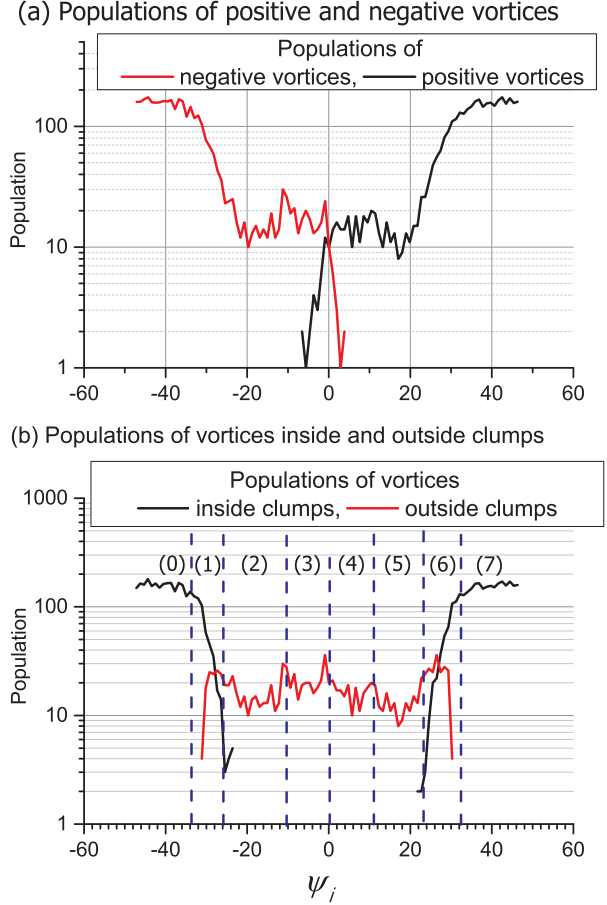}}
  \caption{Normalized population is plotted as a function of energy possessed by 
   each point vortex $\psi_i$. The black and red lines correspond to
  the positive and the negative vortices in (a), and  vortices inside and outside
  clumps in (b), respectively.}
  \label{fig:fig02}
\end{figure}

We observe a system temperature from the quasi-stationary distribution of 
the vortices by the following way.
Let us introduce a notation $N(E)$ that represents the number of vortices 
whose energy $H_i$ satisfies the relation
\begin{equation}
	E \le H_i < E+\Delta E. \label{eqn:inequality}
\end{equation}
Value of $N(E)$ is determined by the vortex distribution.
On the other hand, $N(E)$ is also defined by
\begin{equation}
	N(E) = \left\{
\begin{array}{ll}
	\int_{D(E)} \frac{\omega_{+}(\vr,t)}{\Omega} d\vr, & \mbox{(for positive vortices)}\\
	\int_{D(E)} \frac{\omega_{-}(\vr,t)}{-\Omega} d\vr, & \mbox{(for negative vortices)}\\
\end{array}\right.
\label{eq0}
\end{equation}
where a region in which the inequality (\ref{eqn:inequality})
is satisfied is denoted by $D(E)$.
From now on, we will denote two Eqs. for the positive and negative vortices
into a single formula with the double-sign.
Namely, Eqs. (\ref{eq0}) look like
\begin{equation}
	N(E) = 	\int_{D(E)} \frac{\omega_{\pm}(\vr,t)}{\pm\Omega} d\vr. \label{neweq0}
\end{equation}
The value of the stream function 
satisfying the inequality (\ref{eqn:inequality}) will be denoted by $\Psi$,
\begin{equation}
	\Psi \approx \frac{2}{\pm \Omega}E
\end{equation}
where Eq. (\ref{eqn:Hi}) is used.
During the slow relaxation ($T=100 \sim 200$), vorticity
$\omega$ is a function of the stream function $\psi$ \cite{Yatsuyanagi2015b}
\begin{equation}
	\omega_{\pm} = \omega_{\pm}(\psi).
\end{equation}
Inside the region $D(E)$, vorticity $\omega_{\pm} = \omega_{\pm}(\Psi)$
is approximately constant.
Inserting this into Eq. (\ref{neweq0}), we obtain
\begin{eqnarray}
	N(E) & = & \int_{D(E)} \frac{\omega_{\pm}(\Psi)}{\pm\Omega} d\vr \nonumber\\
	& = & \frac{\omega_{\pm}(\Psi)}{\pm\Omega} \int_{D(E)} d\vr. \label{eq2}
\end{eqnarray}
The integral in the last formula gives the area of the region $D(E)$.
Inserting the Boltzmann-type equilibrium distribution
\begin{equation}
	\omega_{\pm}(\Psi) = \omega_{0\pm} \exp(\mp \beta \Omega \Psi) \label{eq3}
\end{equation}
into Eq. (\ref{eq2}), we finally obtain
\begin{equation}
	\frac{N(E)}{\int_{D(E)} d\vr} = \frac{\omega_{0\pm}}{\pm \Omega} \exp(\mp \beta \Omega \Psi).\label{eq4}
\end{equation}
Equation (\ref{eq4}) states that the population of the vortices $N(E)$ is 
proportional to $\exp(\mp \beta \Omega \Psi)$.
Namely, the temperature can be determined by the population as
the function of energy of each point vortex $\Psi = \psi_i$.
We call $N(E)/\int_{D(E)} d\vr$ a normalized population.
The normalized population is plotted as the function of $\psi_i$ in Fig. \ref{fig:fig02}.
In Fig. \ref{fig:fig02} (a), there are two peaks
at the left and the right ends.
The right peak corresponds to the positive clump near the upper right
corner in Fig. \ref{fig:fig01} at $T=200$, 
as well as the left peak corresponds to the negative clump near the lower left corner.
To confirm the origin of these two peaks,
the population is recalculated separately for the point vortices
inside and outside the clumps in Fig. \ref{fig:fig02} (b).
In the figure, the black line indicates the population of the vortices inside the clumps,
and the red line outside the clumps, which clearly elucidate that the origin of the 
two peaks is due to the clump distribution,
as the dense distribution of the vortices yields high energy vortices.

We categorize the values of $\psi_i$ in 8 parts, (0) through (7) as shown in Fig. \ref{fig:fig02} (b).
Let us first focus on the right black line for the positive vortices.
When the system reaches a thermal equilibrium state, the population
should be proportional to $\exp(-\beta \Omega \psi_i)$.
In region (6), the black line is almost straight with nonzero slope
and expected to be proportional to $\exp(-\beta \Omega \psi_i)$.
We conclude that this slope corresponds to the system temperature as
the slope is almost kept constant among the various simulations
with $N \Omega_0=\mbox{constant}$.
The sign of the slope suggests that $\beta < 0$.
The figure also clearly indicates that the system consists of several subsystems
with different temperature.
In other words, the system still stays in a slow relaxation phase,
which agrees with our conjecture in the previous paper \cite{Yatsuyanagi2015b}.
The same conclusion is drawn from the left black line for the negative vortices.
On the other hand, the slopes of the populations of the vortices in the center of the clump (region
(0) and (7)) and outside the clump (background, region (2) through (5)) are
almost constant which suggests $\beta = 0$.
This is due to the distribution of the vortices has no remarkable structure.

\begin{figure}[ht]
  \resizebox{8cm}{!}{\includegraphics{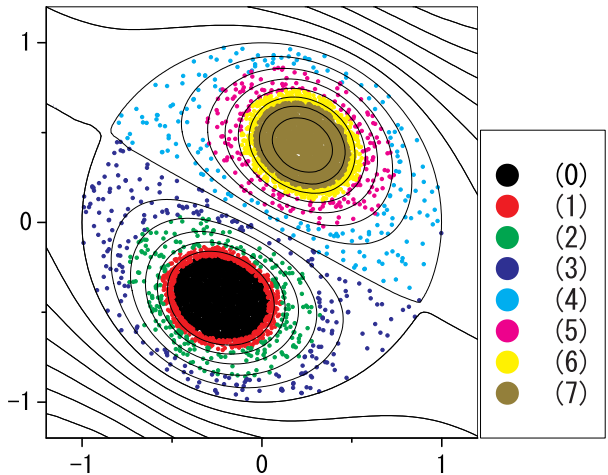}}
  \caption{The distribution is color-coded according to the energy of each point vortex $\psi_i$.}
  \label{fig:fig03}
\end{figure}

Following the arrangements (0) through (7) in Fig. \ref{fig:fig02} (b), 
the distribution in the quasi-stationary state 
shown in Fig. \ref{fig:fig01} at $T=200$ is color-coded. 
The result is shown in Fig. \ref{fig:fig03}.
The black lines in Fig. \ref{fig:fig03} indicate the isosurface of the stream function.
As there is no (macroscopic) flow across the stream line, the clumps sustain their macroscopic 
shapes.
The colors correspond to the densities of the vortices, in other words, the vorticity.
Thus the profile in which the boundary of the regions with different color is parallel
to the stream line directly indicates the system still remains in the slow relaxation 
as $\grad \wz$ is perpendicular to $\vu$.

\begin{figure}[ht]
  \resizebox{6cm}{!}{\includegraphics{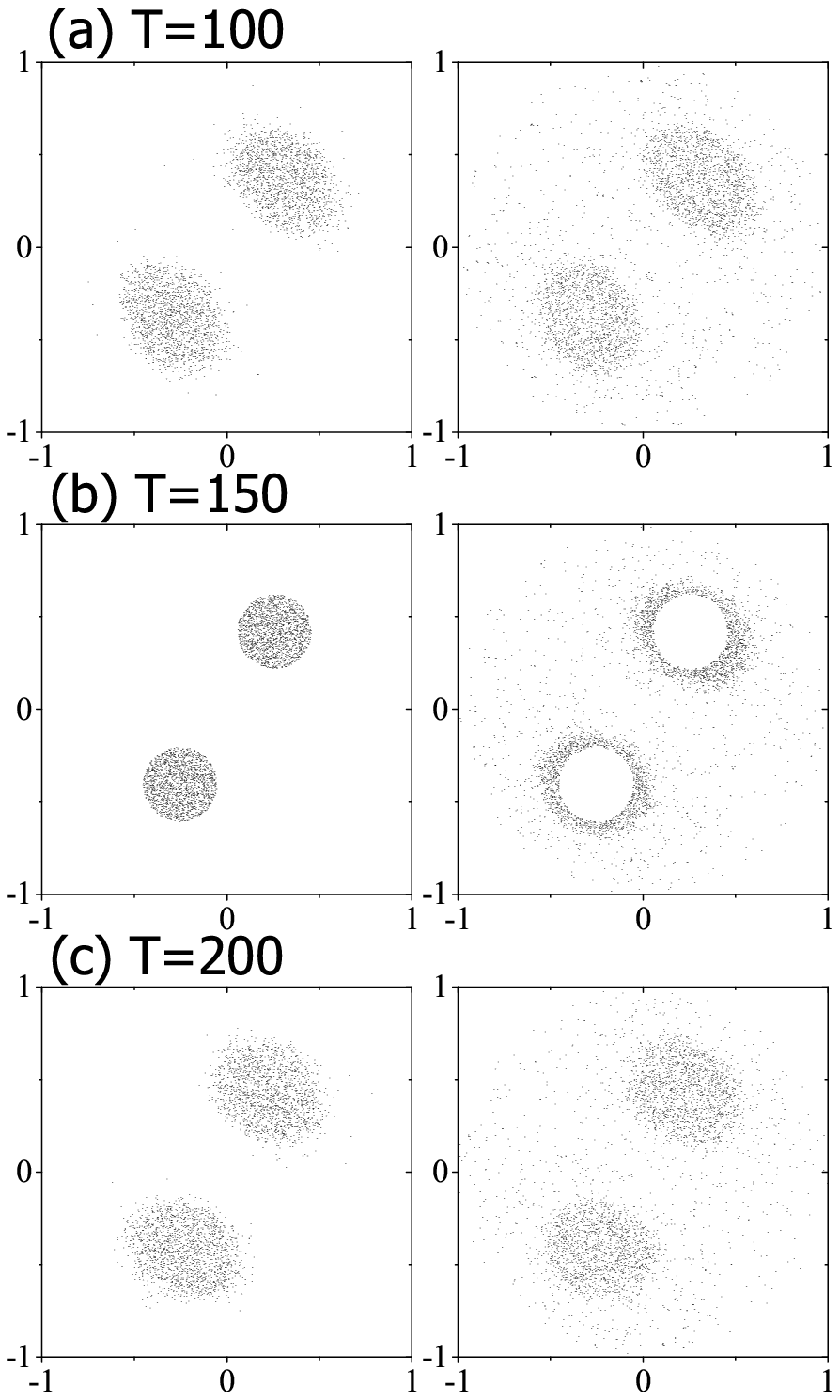}}
  \caption{The effect of the drift term to accumulate the vortices against the diffusion term is demonstrated.}
  \label{fig:fig04}
\end{figure}

We have analytically demonstrated the remarkable feature of the drift term 
appearing in the kinetic equation for the point vortex system \cite{Yatsuyanagi2015b}.
The term acts to accumulate the vortices against the diffusion term.
To check this feature of the drift term numerically, we prepare a result shown in Fig. \ref{fig:fig04}.
At $T=150$ the vortices are labelled following the position, inside or outside the clumps.
Vortices whose distance from the center of gravity of each clump is below a certain value
are chosen as the inside vortices.
In Fig. \ref{fig:fig04}, the left plots in (a), (b) and (c) show the vortices labelled
inside at $T=150$.
In the same way, the right plots show the vortices labelled outside at $T=150$.
The right plots clearly demonstrate the vortices outside the clumps at $T=150$
are transported toward inside the clumps across the stream lines.
This phenomenon can be understood by transforming the kinetic equation like
\begin{equation}
	\frac{\partial \wz}{\partial t} + \div [(\vec{u} + \vec{V})\wz] = -\div (-{\sf D} \cdot \grad \wz) \label{eqn:diff-and-drift}
\end{equation}
where ${\sf D}$ and $\vec{V}$ are the diffusion tensor and the drift velocity, respectively \cite{Yatsuyanagi2015b}.
Equation (\ref{eqn:diff-and-drift}) explicitly states that the vortices are transported by the 
drift velocity $\vec{V}$ even though the macroscopic fluid velocity $\vec{u}$ does not have a 
component perpendicular to the isosurface of the stream function.
It is physically reasonable that the temperature is observable at region (1) and (6)
in Fig. \ref{fig:fig03} where the vortices are transported frequently.

\begin{figure}[ht]
  \resizebox{8cm}{!}{\includegraphics{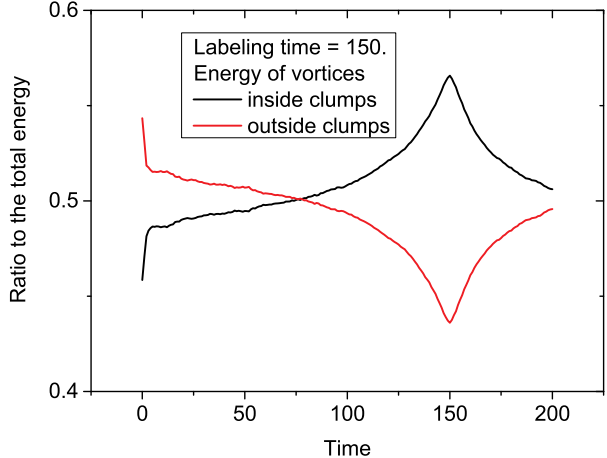}}
  \caption{Sums of the energy of each point vortex are plotted separately for the vortices 
  inside and outside the clumps labelled at $T=150$.}
  \label{fig:fig05}
\end{figure}

Another evidence of the drift term is given in Fig. \ref{fig:fig05}.
The figure shows that the temporal change of the energies of the vortices labelled 
inside and outside at $T=150$.
It visualizes the amount of energy microscopically transferred across the
isosurface of the stream function.
Initially, the distribution of the vortices is rapidly deformed
by the violent relaxation.
Then, the slow relaxation starts and the drift term aggregates the 
vortices toward $T=150$ when the vortices are labelled 
according to the locations.
At that time, energy of the inside vortices maximizes.
After $T=150$, a part of the vortices inside the clumps remain
inside by the drift term and the others are diffused
and transported outside the clumps.
Subsequently, the energy of the inside vortices at $T=150$ lowers.
The number of the vortices inside the clumps is $3198$
out of 6940 (46.1\%) and they occupy 56.5 \% of the total system 
energy at $T=150$.
As they occupy 45.8 \% of the system energy at initial, 
it clearly shows that the energy belonging to the vortices inside 
the clumps at $T=150$ increases due
to the drift term.
Namely, Fig. \ref{fig:fig05} elucidates the energy transportation
along with the microscopic particle transportation across the 
isosurface of the stream function.
This result agrees with the analytical result that the drift term
temporally increases the system energy, while the diffusion term
decreases it with total system energy kept constant \cite{Yatsuyanagi2015b}.

In summary, we have numerically evidenced the drift term
which appears in the kinetic equation for the point vortex system.
The system temperature is determined by the population of the 
vortices as the function of energy of each point vortex $\psi_i$.
The slope of the population indicates the temperature is
negative.
It is physically reasonable that the system temperature is 
characterized by the vortices at the outline 
of the clumps where the vortices are frequently transported
by the diffusion and the drift terms.
The labeling as inside or outside the clumps clearly
demonstrates the vortex transportation by the drift term.
The drift term accumulates the vortices toward the center
of the clumps against the diffusion term.
During the slow relaxation, the vortices are transported across
the isosurface of the stream function, while the macroscopic 
shape of the clumps is temporally unchanged.
We conclude that the self-organization (clump formation) of the point vortex system 
at negative temperature is driven by the drift term.

\end{document}